%% file: main.tex
\documentclass[sigconf]{acmart}
\acmConference[ASE 2022]{IEEE/ACM International Conference on Automated Software Engineering}{10 - 14 October 2022}{Ann Arbor, Michigan, United States}

\usepackage{graphicx}
\usepackage{xspace}
\usepackage{enumitem}
\usepackage[ruled,vlined,linesnumbered]{algorithm2e}
\usepackage[capitalise]{cleveref}
\usepackage{algpseudocode}
\usepackage{subfig}
\usepackage{amsmath}
\usepackage{multirow}
\usepackage{setspace}
\usepackage{listings}
\usepackage{xcolor}
\usepackage{amsmath}
\usepackage{diagbox}
\usepackage{tabularx}
\usepackage{booktabs}
\usepackage{url}
\usepackage{textcomp}
\usepackage{caption}
\usepackage{balance}
\usepackage{pifont}

\Crefname{lstlisting}{Listing}{Listing}
\lstdefinelanguage{JavaScript}{
  morekeywords=[1]{break, continue, delete, else, for, function, if, in,
    new, return, this, typeof, var, void, while, with},
  morekeywords=[2]{false, null, true, boolean, number, undefined,
    Array, Boolean, Date, Math, Number, String, Object},
  morekeywords=[3]{eval, parseInt, parseFloat, escape, unescape},
  sensitive,
  morecomment=[s]{/*}{*/},
  morecomment=[l]//,
  morecomment=[s]{/**}{*/}, 
  morestring=[b]',
  morestring=[b]"
}[keywords, comments, strings]
\AtBeginDocument{%
  \providecommand\BibTeX{{%
    \normalfont B\kern-0.5em{\scshape i\kern-0.25em b}\kern-0.8em\TeX}}}

\setcopyright{acmcopyright}
\copyrightyear{2022}
\acmYear{2022}
\setcopyright{acmcopyright}\acmConference[ASE '22]{37th IEEE/ACM International
Conference on Automated Software Engineering}{October 10--14, 2022}{Rochester,
MI, USA}
\acmBooktitle{37th IEEE/ACM International Conference on Automated Software
Engineering (ASE '22), October 10--14, 2022, Rochester, MI, USA}
\acmPrice{15.00}
\acmDOI{10.1145/3551349.3560427}
\acmISBN{978-1-4503-9475-8/22/10}

\usepackage[font=small,skip=3pt]{caption}
\newcommand{\distance}{3pt}
\newcommand{\revised}[1]{{\textcolor{black}{#1}}}

\newcommand{\target}{JavaScript-based\xspace}

\begin{document}

\title{Towards Understanding the Faults of JavaScript-Based \\Deep Learning Systems}
\renewcommand{\shorttitle}{Towards Understanding the Faults of JavaScript-Based Deep Learning Systems}

\author{Lili Quan}
\affiliation{
    \institution{College of Intelligence and Computing, Tianjin University}
    \city{Tianjin}
    \country{China}
 }

\author{Qianyu Guo}
\affiliation{
 \institution{Zhongguancun Laboratory}
 \city{Beijing}
 \country{China}
 }

\author{Xiaofei Xie}
\affiliation{%
  \institution{Singapore Management University}
  \country{Singapore}
}

\author{Sen Chen}
\authornote{Sen Chen (senchen@tju.edu.cn) and Xiaohong Li (xiaohongli@tju.edu.cn) are the corresponding authors.}
\affiliation{
    \institution{College of Intelligence and Computing, Tianjin University}
    \city{Tianjin}
    \country{China}
 }

\author{Xiaohong Li}
\authornotemark[1]
\affiliation{
    \institution{College of Intelligence and Computing, Tianjin University}
    \city{Tianjin}
    \country{China}
 }

\author{Yang Liu}
\affiliation{
 \institution{Nanyang Technological University}
 \country{Singapore}
}

\renewcommand{\shortauthors}{Lili Quan and Qianyu Guo, et al.}

\begin{abstract}
    \input{0-abstract}
\end{abstract}


\maketitle
\input{1-introduction}
\input{2-approach}
\input{3-symptom}

\input{4-rootcause}
\input{5-fixpattern}
\input{6-difference}

\input{7-discussion}
\balance
\input{8-related-work}
\input{9-conclusion}

\begin{acks}
\input{10-ack}
\end{acks}

\clearpage

\bibliographystyle{ACM-Reference-Format}
\bibliography{reference}

\end{document}

%% file: 0-abstract.tex
Quality assurance is of great importance for deep learning (DL) systems, especially when they are applied in safety-critical applications. 
While quality issues of native DL applications have been extensively analyzed, the issues of \target DL applications have never been systematically studied. Compared with native DL applications, \target DL applications can run on major browsers, making the platform- and device-independent. 
Specifically, the quality of \target DL applications depends on the 3 parts: the application, the third-party DL library used and the underlying DL framework (e.g., TensorFlow.js), called \target DL system.
In this paper, we conduct the first empirical study on the quality issues of \target DL systems. 
Specifically, we collect and analyze 700 real-world faults from relevant GitHub repositories, including the official TensorFlow.js repository, 13 third-party DL libraries, and 58 \target DL applications.
To better understand the characteristics of these faults,
we manually analyze and construct taxonomies for the fault symptoms, root causes, and fix patterns, respectively. Moreover, we also study the fault distributions of symptoms and root causes, in terms of the different stages of the development lifecycle, the 3-level architecture in the DL system, and the 4 major components of TensorFlow.js framework.
Based on the results, we suggest actionable implications and research avenues that can potentially facilitate the development, testing, and debugging of \target DL systems.

%% file: 1-introduction.tex
\section{Introduction}\label{sec:intro}
Deep learning (DL) has been widely applied into various applications such as image classification~\cite{resnet}, natural language processing~\cite{wu2016google}, and speech recognition~\cite{graves2013speech}. To support the DL-based applications, many DL libraries and frameworks such as TensorFlow~\cite{tensorflow}, PyTorch~\cite{paszke2017automatic}, and Keras~\cite{keras} have been developed and widely used. However, DL systems have been demonstrated to be vulnerable (e.g., adversarial attack~\cite{goodfellow6572explaining,chen2021real,chen2022as2t,chen2018automated}), which can cause serious consequences when they are applied to some safety-critical applications such as healthcare~\cite{liu2014early} and autonomous driving~\cite{chen2015deepdriving}. Hence, quality assurance of DL systems is required. 

Recently, extensive researches have been conducted from various communities including AI, software engineering, and security to study the quality issues of DL systems. 
For example, a lot of adversarial attack techniques~\cite{fgsm,papernot2016practical,brendel2017decision} have been proposed to evaluate the model robustness. The quality of DL frameworks is also important for DL systems. 
Some works including bug analysis~\cite{islam2019comprehensive,chen2021empirical,xiao2018security,humbatova2020taxonomy} and framework testing~\cite{pham2019cradle,guo2020audee,wang2020deep} have been studied for DL frameworks. In addition to the model and DL frameworks, some studies are conducted on the programming bugs of DL applications (e.g., programming bugs with TensorFlow~\cite{zhang2018empirical}, bugs on the model architectures~\cite{zhang2021autotrainer}). However, most of the studies focus on the native applications that can run on specific environments (e.g., Android and iOS) and DL frameworks.

\begin{figure}
    \centering
    \includegraphics[width=0.34\paperwidth]{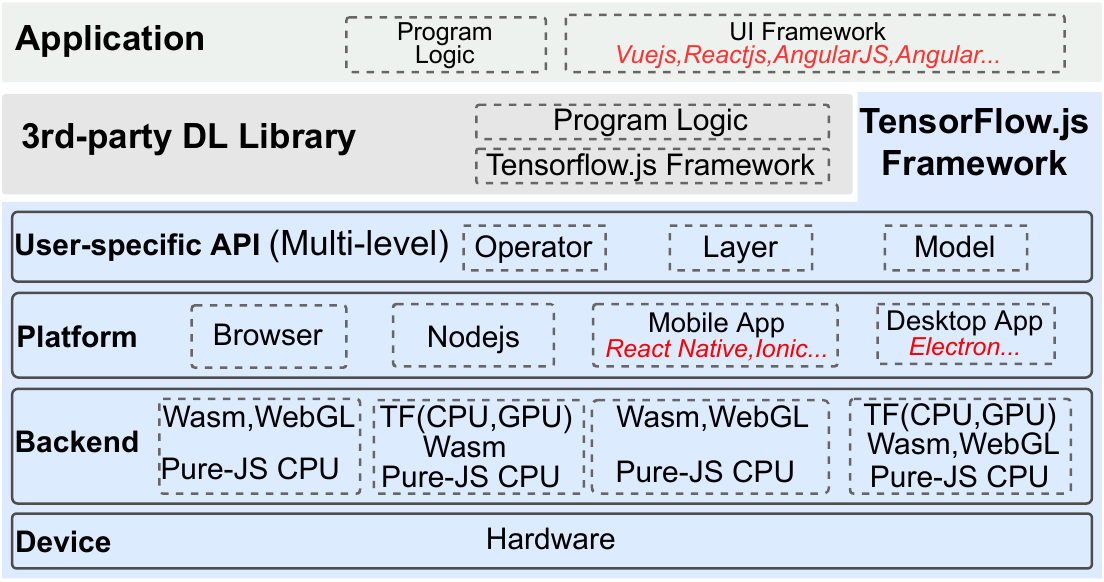}
    \caption{The typical architecture of \target DL system}
    \label{fig:architecture-of-jsdl-system}
\end{figure}

A main drawback of the native applications is that they are often platform-specific (e.g., Windows, iOS, and Android) and device-specific (e.g., PC, mobile phones, and IoT devices). Considering the requirements for easy deployment and migration, \target DL applications are becoming more and more popular. Compared to native DL applications, \target applications are platform-agnostic and device-agnostic because they can easily run on major browsers such as Chrome, Firefox, and Safari on different platforms and devices~\cite{ma2019moving}. Various JavaScript-based DL frameworks and libraries (e.g., TensorFlow.js~\cite{TensorFlow.js}, Keras.js~\cite{kerasjs}, and ML5.js~\cite{ml5js}) have also been developed. \revised{While the quality issues of native applications have received a lot of attention, the quality study of \target DL applications is still less touched, which also motivates 
this work to study the faults in JavaScript-based DL systems and their unique characteristics compared with native applications.}

To fill this knowledge gap, in this paper, we conduct the first empirical study towards understanding the faults in JavaScript-based DL applications that run on multiple platforms (i.e., browsers, Node.js, mobile apps, and desktop apps).
The quality of a JavaScript-based DL application typically depends on three parts: the application itself, the 3rd-party DL libraries used in the application and the underlying \target DL framework. As shown in~\Cref{fig:architecture-of-jsdl-system}, DL software including the 3-level architecture (i.e., application level, 3rd-party DL library level, and framework level) is called the \target DL system. The faults in any of the three levels can significantly affect the quality of the entire system. Hence, in this paper, we conduct a comprehensive study on faults of all the \textit{three levels}. In particular, for the DL framework, this paper focuses on TensorFlow.js which is the most popular JavaScript-based DL framework. As shown in~\Cref{fig:architecture-of-jsdl-system}, TensorFlow.js contains \textit{four major components} (i.e., API, Platform, Backend, and Device).
\revised{Note that it contains several DL backends (e.g., WebGL and Wasm) specific to JavaScript, compared to native DL frameworks (e.g., TensorFlow).}

\begin{figure}
    \centering
    \includegraphics[width=0.35\paperwidth]{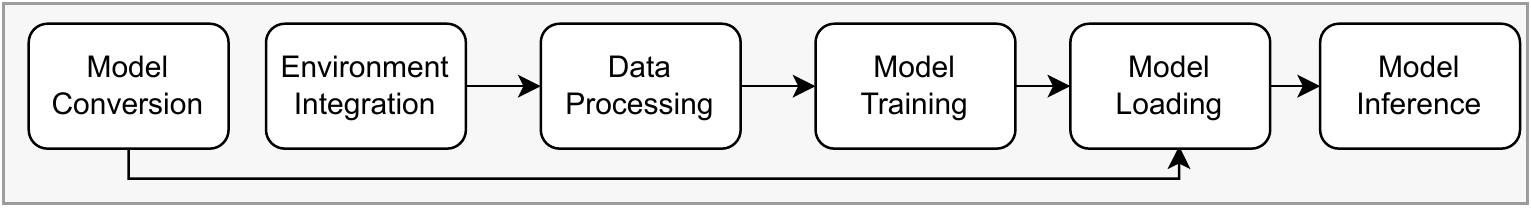}
    \caption{The typical developing stage of \target DL system}
    \label{fig:developing-stage}
\end{figure}

On the other hand, we can observe the faults from the development lifecycle of \target DL systems.
As shown in~\Cref{fig:developing-stage}, the lifecycle usually includes \textit{6 stages}~\cite{islam2019comprehensive,du2020fairness,chen2022toward,hapke2020building}. 
Specifically, \textit{Environment Integration} refers to integrating DL framework (i.e., TensorFlow.js) into the applications. \textit{Data Processing} mainly focuses on preprocessing the input and post-processing the model inference results. \textit{Model Training} aims to build and train the model. \textit{Model Conversion}
{converts models taken from other platforms into the target format}. \textit{Model Loading} loads the models through relevant APIs. \textit{Model Inference} performs the prediction.

Specifically, we collected {72} relevant GitHub repositories including 1 official TensorFlow.js, 13 3-party DL libraries, and 58 \target DL applications that cover the 3-level architecture shown in \Cref{fig:architecture-of-jsdl-system}. We collected 700 faults in total from these repositories. Based on these 700 faults, we perform a comprehensive analysis to investigate their symptoms, root causes, and fix patterns. We also highlight the unique characteristics of \target DL systems compared to the bugs of native applications. 

From these 700 faults, we summarized {26} symptoms, {17} root causes, and {16} fix patterns. Furthermore, we study the distribution of symptoms on the \textit{6 stages} of the developing lifecycle; the distribution of root causes on the \textit{3-level} architecture, and \textit{4 components} of TensorFlow.js. 
The classification results (i.e., symptoms, root causes, and fix patterns) and the distribution results can help developers and researchers better understand the various faults and their characteristics, providing insights for developing different testing, debugging, and repairing techniques on \target DL systems.

In summary, we make the following main contributions:
\begin{itemize}[leftmargin=*]
    \item To the best of our knowledge, this is the first empirical study towards understanding the characteristics of the faults in \target DL systems. We constructed the taxonomies for fault symptoms, root causes, and fix patterns respectively, \revised{and further discussed the different characteristic of faults between the native DL systems and the \target DL systems.}
    \item We studied the fault distributions of symptoms and root causes on the 6 stages of the lifecycle of DL system, the 3-level architecture in the DL system, and the 4 components of TensorFlow.js.
    \item We provided a series of findings that benefit multiple stakeholders such as application developers, 3rd-party DL library developers, framework developers, and researchers in \target DL ecosystems.  
    \item We collected a dataset of real faults from a wide spectrum of sources, including the official TensorFlow.js repository, the 3rd-party DL libraries based on TensorFlow.js, and the high-level applications, which can be a valuable benchmark for further analyzing and testing the \target DL ecosystems.
    We have made the fault dataset publicly available to facilitate the new research field. More details can be found on our website~\cite{website}.
\end{itemize}

%% file: 2-approach.tex
\section{Empirical Study Methodology}\label{sec:approach}

\subsection{Study Design}

\begin{figure*}
    \centering
    \includegraphics[width=0.6\paperwidth, height=3.65cm]{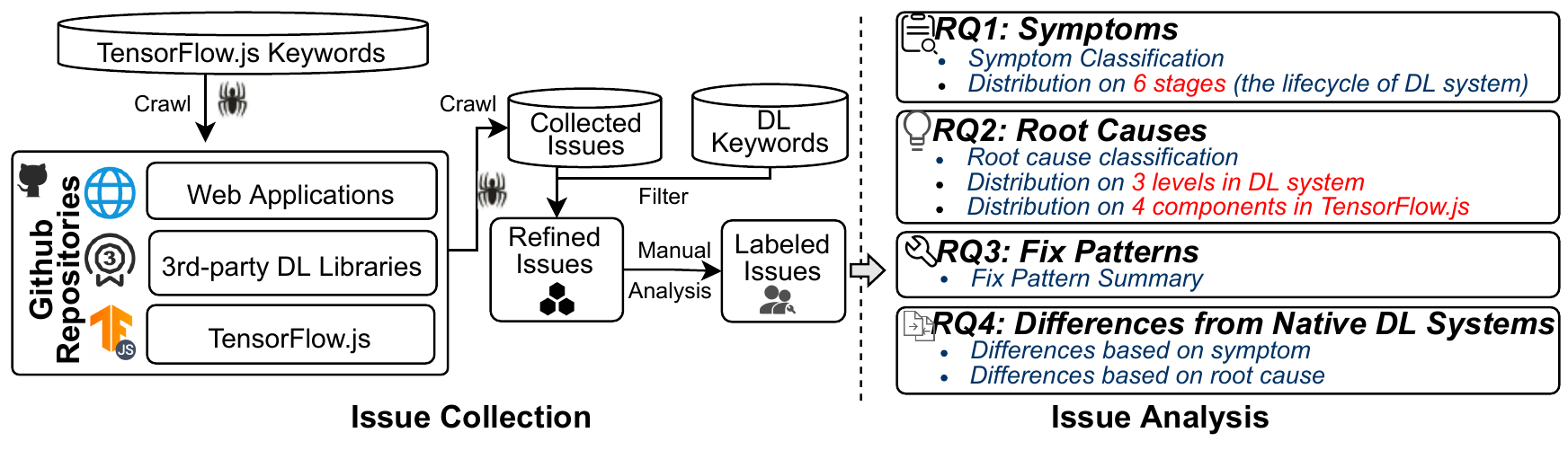}
    \vspace{-2mm}
    \caption{Overview of our methodology}
    \label{fig:overview}
\end{figure*}

To characterize issues in \target DL systems, we first collect and analyze relevant repositories from GitHub. As \target DL systems can be built on top of various \target DL frameworks, in this work, we mainly focus on the DL systems developed with TensorFlow.js~\cite{TensorFlow.js}, which is the most popular \target DL framework. The overview of the methodology is illustrated in \Cref{fig:overview}. 
We first collect popular Github repositories through keyword search, including the official TensorFlow.js repository, the 3rd-party DL library repositories that wrap TensorFlow.js, and the repositories of DL-based web applications based on TensorFlow.js. For each repository, we crawl issues that may be related to fixing/discussing relevant problems and construct the candidate dataset for further analysis.

With the labeled issues, {we study 4 research questions (i.e., the symptoms, root causes, fix patterns, and the differences from native DL systems}).
For the analysis of the symptoms in \textbf{RQ1}, we first summarize the taxonomy of fault symptoms and then analyze the distribution of symptoms on the 6 stages involved in the development of \target DL systems. The findings from \textbf{RQ1} can provide insights for understanding and detecting bugs for \target DL systems.
The root cause analysis in \textbf{RQ2} aims to characterize the fundamental reasons for these faults. We first summarize the different types of root causes, and further analyze the distribution of these root causes on the 3-level architecture and the components of TensorFlow.js, respectively. 
We summarize the \textit{fix pattern} in \textbf{RQ3} aiming to characterize the solutions to fix these faults.
Finally, in \textbf{RQ4}, we analyze the different features of the fault symptoms and fault root causes between the native DL systems studied in the previous work and JavaScript-based DL systems.

\subsection{Data Collection}\label{sec: Data Collection}

Following existing work~\cite{islam2019comprehensive,humbatova2020taxonomy,chen2021empirical,gu2021demystifying,chen2022toward}, we first use the GitHub search API~\cite{GithubSerachAPI} to collect repositories that are related to \target DL systems, including the DL framework TensorFlow.js~\cite{TensorFlow.js}, 3rd-party DL libraries, and web applications using DL. In HTML and JavaScript code, TensorFlow.js is usually imported using the \texttt{script tag}\footnote{<script src="https://cdn.jsdelivr.net/npm/@tensorflow/tfjs/dist/tf.min.js"> </script>
} and the statement \texttt{import * as tf from @tensorflow/tfjs} respectively. Therefore, we use ``tfjs'' and ``TensorFlow.js'' as the keywords to search the repositories. For each repository, we also collect the attributes such as links, number of stars~\cite{Githubstar}, number of forks~\cite{Githubfork}, number of issues, and language type. In total, we collected 924 candidate repositories. We filter out non-JavaScript-based and unpopular repositories based on the following criteria: \textbf{1)} the language type is not HTML, JavaScript or TypeScript~\cite{TypeScript}; \textbf{2)} there are no issues; and \textbf{3)} the total number of stars and forks is less than 10. In addition, we manually check the remaining repositories to exclude irrelevant repositories that are not real DL systems, e.g., some tutorials, books, or repositories that contain the keyword but do not actually use the TensorFlow.js. 
In the end, 72 repositories are selected, including 1 official TensorFlow.js framework, 13 3rd-party DL libraries, and 58 web applications. The details of the repositories can be found on our website~\cite{website}.

Based on the repositories, we then collect issues before Dec. 2021 for the following study because these issues contain more detailed information such as original reports, discussions between users and developers, and the fix strategy. \cref{tab: crawledissuenum} shows the number of issues collected under each type of repository, where AF and MF stand for \textit{Automatic Filtering} and \textit{Manual Filtering}, respectively.
A total of 3,859 issues are crawled initially, of which 2,374 are from the official TensorFlow.js repository, 1,194 are from 13 3rd-party DL libraries, and 291 are from 58 web applications. We exclude the issues with the corresponding label (e.g., \textit{stat:awaiting response}) or without answers. 
Note that, considering that manually analyzing bugs is time-consuming, we cannot analyze all the historical issues of TensorFlow.js. Therefore, to balance scale and cost, we filter out the issues related to deprecated versions of TensorFlow.js (before 2020-01-01).

For issues from 3rd-party DL libraries and web applications, we adopt the similar filtering strategies used in previous work~\cite{humbatova2020taxonomy} to discard the DL irrelevant issues. Specifically, in~\cite{humbatova2020taxonomy}, a vocabulary of relevant words (e.g., ``epoch'') related to native DL frameworks (e.g., TensorFlow) are defined, and all issues without these words are excluded. Considering the difference between TensorFlow.js and native DL frameworks, we update the vocabulary with TensorFlow.js-specific keywords (e.g., dispose, WebGL). The vocabulary finally contains 147 relevant words. As a result, 1,293 refined issues are selected as shown in column \textit{After AF} of \cref{tab: crawledissuenum}. Furthermore, during the manual labeling process (see \cref{sec: Manual Labeling}), we discard issues with unclear descriptions and false positives. 
For example, some issues contain the keywords but are not errors. 
We finally obtain 684 issues, of which 359, 291, and 34 are from the TensorFlow.js, 3rd-party DL libraries, and web applications, respectively.

\begin{table}[]
\centering
\caption{TensorFlow.js-related issues on GitHub}
\label{tab:issue_dataset}
\scalebox{0.75}{\begin{tabular}{@{}l|c|r|r|r@{}}
\hline
\textbf{Repository Type} & \textbf{\#Repositories} & \textbf{\#Issues Crawled} & \textbf{After AF} & \textbf{After MF} \\ \hline
\textbf{Official TensorFlow.js}& 1 & 2,374 & 463 & 359 \\
\textbf{3rd-Party DL Library} & 13 & 1,194 & 724 & 291 \\
\textbf{Web Applications} &58 & 291 & 106 & 34 \\\hline
\textbf{Total} & 72 & \textbf{3,859} & \textbf{1,293} & \textbf{684} \\ \hline
\end{tabular}}
\label{tab: crawledissuenum}
\end{table}

\subsection{Manual Labeling}\label{sec: Manual Labeling}

To answer the research questions,
we manually label the faults in the 684 issues from 6 aspects: (1) \textit{symptoms}, which show what the fault looks like, (2) \textit{development stages},
which show at which stage the error happens, (3) \textit{root causes}, which explain why the faults occur, (4) \textit{3-level architecture}, showing at which level of the DL system a root cause comes from, (5) \textit{components of TensorFlow.js}, indicating which component the root cause of a framework-related failure came from and (6) \textit{fix patterns}, which describes how a fault is resolved. 
Note that we need to construct the labels for the \textit{symptoms}, \textit{root causes} and the \textit{fix patterns} in our study. The labels about \textit{development stages}, \textit{3-level architecture} and \textit{components of TensorFlow.js} are fixed (see details in ~\Cref{fig:architecture-of-jsdl-system}), which are used to perform the distribution analysis of the symptoms and the root causes.
The classification and the distribution analysis can help researchers and developers better understand, detect and fix different kinds of faults.

Regarding the labeling, we first randomly sample 50\% of issues for pilot labeling. The first two authors label each fault of the issue following an open coding procedure~\cite{seaman1999qualitative}. 
Specifically, they carefully read each issue's title, descriptions and inter-developer discussions to understand the context, and construct the taxonomies for symptoms, root causes and fix patterns by grouping similar faults together into categories. The taxonomies are adjusted continuously in the construction process. 
During the labeling process, any disagreement is resolved by an arbitrator, who has more than five years of experience in DL-relevant research. All labels and taxonomies are finally discussed and finalized by all participants.

Second, the first two authors independently label the faults in the remaining issues based on the taxonomies generated in the pilot labeling. The issues that cannot be classified into the current taxonomies are labeled with a new category. The labeling process involves five rounds, and 20\% of the remaining issues are labeled in each round. Following existing work~\cite{chen2022toward,gu2021demystifying}, we adopt the Cohen’s Kappa coefficient~\cite{cohen1960coefficient} to measure the inter-rater agreement of the independent labeling. After the first round, the Cohen’s Kappa coefficient is just about 39\% and then the two authors discuss all these inconsistent results with the arbitrator.  After the second round, the Cohen’s Kappa coefficient reached 70\%. 
Through further discussion with the arbitrator on inconsistencies, the Cohen’s Kappa coefficients are over 90\% after all the subsequent rounds.
\revised{After manual labeling, we identify 700 faults from 684 issues collected, of which 16 issues contain 2 faults.}

%% file: 3-symptom.tex
\section{Symptoms (RQ1)}\label{sec:Symptom}
\subsection{Symptom Classification Results \label{sec:symptom-classification}}
\begin{figure*}
    \centering
    \includegraphics[width=0.7\paperwidth]{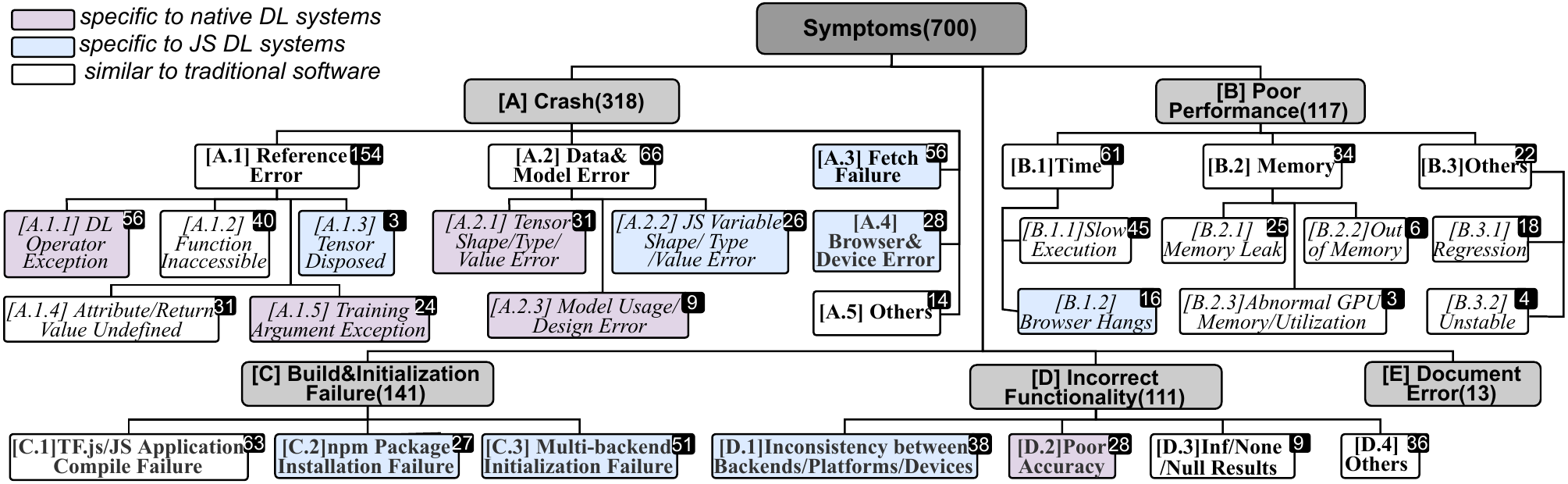}
     \vspace{-1mm}
    \revised{\caption{Symptom taxonomy of faults in \target DL systems
    \label{fig:symptom-classification}}
    }
\end{figure*}

\Cref{fig:symptom-classification} shows the hierarchical taxonomy of fault symptoms in \target DL systems.
It is grouped into 5 high-level categories (i.e., \textit{Crash}, \textit{Build \& Initialization Failure}, \textit{Poor Performance}, \textit{Incorrect Functionality}, and \textit{Document Error}), 15 inner categories, and 15 leaf categories specific to \textit{Reference Error}, \textit{Data\&Model Error}, and \textit{Poor Performance}.
\revised{Note that the blue and pink rectangles indicate symptoms specific to JavaScript-based and native DL systems respectively, which will be further detailed in~\Cref{sec:symptom difference}.}

\vspace{-1mm}
\subsubsection{\textbf{Crash (A)}}
This category indicates the functionality of DL systems is terminated unexpectedly with error messages like ``undefined'' or ``uncaught Error'', accounting for the largest proportion (45.43\%) of faults in this study, including 318/700 faults and 5 subcategories. \revised{Note that non-functional terminations like performance issues (e.g., out of memory) are not included in this category.} 

Among them, \textit{Fetch Failure (A.3)} and \textit{Browser \& Device Error (A.4)} mainly appear in the browser-based DL tasks, accounting for 26.42\% of all crashes in this study.
\textit{Browser \& Device Error} refers to the crashes showing messages that browsers or devices are problematic. 
For example, WebGL (i.e., a JavaScript API for rendering high-performance graphics on browsers, which can be used to accelerate DL tasks) is not supported on a Macbook Pro 2018~\cite{issues4768}.
\target DL systems need to request model files or data
via the web API (i.e., \texttt{Fetch}~\cite{Fetch_API}). \textit{Fetch Failure} occurs during this process, which refers to crashes with error messages showing the fetch failure, i.e., it cannot directly access the local file system due to the same-origin policy~\cite{Same-origin-policy} that browsers follow.
These 2 subcategories jointly reveal that \target DL systems are affected by the limitations of browsers and devices.

\revised{\textit{Reference Error} is the most common crash type with 154 faults (see A.1 in \Cref{fig:symptom-classification}), referring to that certain objects (i.e., function, variable, and training argument) are not implemented, defined, or found.
Specifically, the function reference errors include DL-related function exceptions (i.e., \textit{A.1.1}) 
and traditional function exceptions (i.e., \textit{A.1.2}).
Variable reference errors refer to disposed tensors and undefined variable properties/function return values are accessed by program (i.e., \textit{A.1.3 and A.1.4}).
The remaining 24 faults are \textit{Training Argument Exception} (\textit{A.1.5}).
Among them, \textit{A.1.1} (DL Operator Exception) is the most common,
indicating the implementations of current \target DL systems are still in fragile status, which can easily bring errors when using a lot of DL functions.}

\revised{The \textit{Data \& Model Error (A.2)} refers to the crashes that data or model is reported to be problematic.
Specifically, data errors represent the incorrect data types, shapes, and values, involving both DL-related tensors (see A.2.1 in \Cref{fig:symptom-classification})
and JavaScript variables (see A.2.2 in \Cref{fig:symptom-classification}).
For example, the invalid data shape reports ``\textit{tensor should have 131072 values but has 14636}''~\cite{issue5821}, and the invalid data type reports ``expected input to be of type HTMLImageElement''~\cite{issues788}.
The model errors usually show failure occurs on model usage or structure construction with messages like ``\textit{model needs to be complied before used}''. }
The remaining 14 crashes with low frequency (occurring only once or twice) are thus categorized in \textit{Others}.

\vskip 1mm
\noindent \fbox{
	\parbox{0.95\linewidth}{\textbf{Finding}: Crash is the most common symptom, accounting for 45.43\% of all faults. In particular, \textit{Fetch Failure} and \textit{Browser \& Device Error} mainly appear when performing DL tasks on the browsers. Meanwhile, the 154 \textit{Reference Error} indicates that the \target DL systems are still in fragile status.}
}

\subsubsection{\textbf{Build \& Initialization Failure (C)}}
\revised{At the start of developing \target DL applications, developers need to build and initialize the necessary environments. 141/700 (20.14\%) faults occurred in this process, which consists of 3 typical symptoms. 
Specifically, 63 faults belong to the building failure when developers compile TensorFlow.js from source code and further compile the DL applications (i.e., \textit{C.1} in~\Cref{fig:symptom-classification}).  
{Alternatively, developers can install compiled TensorFlow.js via NPM (i.e., the package manager for JavaScript), during which 27 faults were found} (\textit{C.2} in Figure 4).}

Apart from the explicit building errors mentioned above, there may still be exceptions even after the application has been successfully compiled. Consider the \textit{Multi-backend Initialization Failure (C.3)}, 51 faults show that JavaScript-based DL systems fail to initialize certain DL backends (e.g., Wasm) even though they have already installed TensorFlow.js successfully~\cite{issue4593}.
The main reasons include: the device/browser used is incompatible with the backend, and some errors in the implementation of TensorFlow.js. More than 20\% of the faults occurring during the building or initialization process indicate the complexity of the \target DL system.

\vskip 1mm
\noindent \fbox{
	\parbox{0.95\linewidth}{	\textbf{Finding}: 
		\revised{20.14\% of all faults are introduced when building and initializing the necessary environments for \target DL systems, which is the second most common symptom.}}
}

\subsubsection{\textbf{Poor Performance (B)}}    
\textit{Poor Performance} is another typical symptom for \target DL systems, which slow down the execution processes, consume excessive resources, and bring bad user experiences. 117/700 faults belong to this category, covering 16.71\% of all faults.
It is organized into 3 inner categories (i.e., \textit{Time}, \textit{Memory}, and \textit{Others}) and 7 leaf categories, as shown in ~\Cref{fig:symptom-classification}.

\textbf{Time (B.1).} This category covers the performance faults exhibiting high time cost, which accounts for the largest portion of \textit{Poor Performance}, i.e., 57.26\%. Particularly, 38.46\% of performance faults show \textit{Slow Execution Time} when performing DL tasks,
including data processing, model building, training, and prediction. The systems can still work but are extremely slow. 
13.68\% of the performance faults result in a more severe symptom (i.e., \textit{Browser Hangs}) that \target DL systems cease to respond to inputs. For example, the desktop browsers hang and cannot respond over a long period of time~\cite{issues826}.

\textbf{Memory (B.2)}. This category covers 29.06\% of the performance faults which consume RAM/GPU memory abnormally. It contains 3 subcategories: the \textit{Memory Leak (B.2.1)}, \textit{Out of Memory (B.2.2)}, and \textit{Abnormal GPU Memory/Utilization (B.2.3)}. Specifically, \textit{B.2.2} is the most severe symptom and can cause the \target DL systems to terminate unexpectedly.
Moreover, \textit{B.2.1} is the most common symptom in this category, which can lead to out of memory in severe cases.
The remaining 3 memory faults show unexpectedly high or low GPU memory usage, i.e., \textit{B.2.3}.

\textbf{Others (B.3).}
We also summarize two special types of performance faults, i.e., the \textit{Regression} and \textit{Unstable}, covering 18.80\% of all performance faults in this study. Specifically, 
\textit{Regression} refers to the faults occurring after the TensorFlow.js upgradation. For example, the fragment shader compilation fails after TenorFlow.js upgrading from version 3.5.0 to 3.6.0~\cite{issues5246}.
\textit{Unstable} means the inference results of \target DL systems are unstable. For example, when the portrait in front of the camera remains still, the face recognition results are constantly changing~\cite{issues59}.

\textit{Poor Performance} accounts for a considerable proportion of all faults in this study, and it can directly affect the user experience.
There are two main reasons for the poor performance: 
1) Web applications inherently suffer from low performance due to the use of DOM~\cite{dom} tree in the browser. 2) The explicit memory management can easily introduce memory performance issues, e.g., manually releasing memory. It is especially required on the WebGL backend, because the browser does not automatically recycle WebGLTextures, a variable where tensor data is ultimately stored.

\vskip 1mm
\noindent \fbox{
	\parbox{0.95\linewidth}{	\textbf{Finding}: As a kind of non-functional fault, \textit{Poor Performance} covers 16.71\% of all faults in this study. It has various symptoms, e.g., more than one-third of the performance faults slow down \target DL systems, and nearly 30\% of the performance faults consume extremely high memory. }
}

\subsubsection{\textbf{Incorrect Functionality (D)}}    

We find another kind of faults that can run normally without crashes/failures, but the final results are incorrect. We refer to these faults as the \textit{Incorrect Functionality}, covering 111/700 (15.86\%) faults in this study.
Specifically, 38 faults show that \target DL systems produce different results under multiple DL backends, platforms, or devices (D.1).
Models may give wrong inference results under some data in 28 faults (known as \textit{Poor Accuracy (D.2)}) and even non-numerical outputs in 9 faults, such as the infinity and Null/None results (\textit{D.3}).
Besides, there are other discrete cases (36 faults) that \target DL systems provide incorrect functionality. For example, TensorFlow.js can not properly switch the WebGL backend to the CPU backend ~\cite{issue5632}.

Notably, we clarify the incorrect functionality involves two levels. 
1) The DL system level. That is, \target DL systems give an incorrect inference result. For example~\cite{issues5486}, when there is a hand/phone in the camera, the \textit{blazeface} infers that there is a face, indicating the inference of the DL system is not robust. 
2) The DL operator level. Namely, the DL operators in TensorFlow.js give wrong calculation results without crashing. For example~\cite{issues5800}, with an input \texttt{NaN}, the operator \texttt{tf.isNaN} outputs \texttt{FALSE}, indicating the implementation of this operator is incorrect.
We emphasize this is a severe symptom that should arouse more attention from TensorFlow.js vendors.
Due to the statistical characteristics of DL models in decision-making, the inference outputs show more uncertainty and uninterpretability than traditional software, which require carefully-designed oracles to capture the unexpected results.   

\noindent \fbox{
	\parbox{0.95\linewidth}{	\textbf{Finding}: \textit{Incorrect Functionality} accounts for 15.86\% of all faults in this study.
		This symptom category appears not only in the DL system but also in the specific operator.
		Moreover, we need to design test oracles for capturing these faults.}
}

\subsubsection{\textbf{Document Error (E)}}
\textit{Document Error} refers to the faults related to TensorFlow.js official documents/tutorials, including invalid links, incorrect instructions, and missing tutorials.  
Although the 13 documental faults only account for 1.86\% in this study, they will not only bring bad experiences to TensorFlow.js users but also may cause implementation bugs or even security vulnerabilities for the entire DL systems. Similar implementation problems caused by poor-quality docs have been extensively studied in other fields~\cite{chen2016host}. As the foundation of DL development in JavaScript ecosystems, the rigorousness and correctness of the TensorFlow.js guidance docs should also be seriously paid attention to. 

\noindent \fbox{
	\parbox{0.95\linewidth}{	\textbf{Finding}: Although the \textit{Document Error} only accounts for a small proportion (i.e., 1.86\%), it will bring bad experiences to the TensorFlow.js users.}
}

\subsection{Symptom Distribution}

\begin{figure}[t!]
    \centering
    \includegraphics[width=0.35\paperwidth]{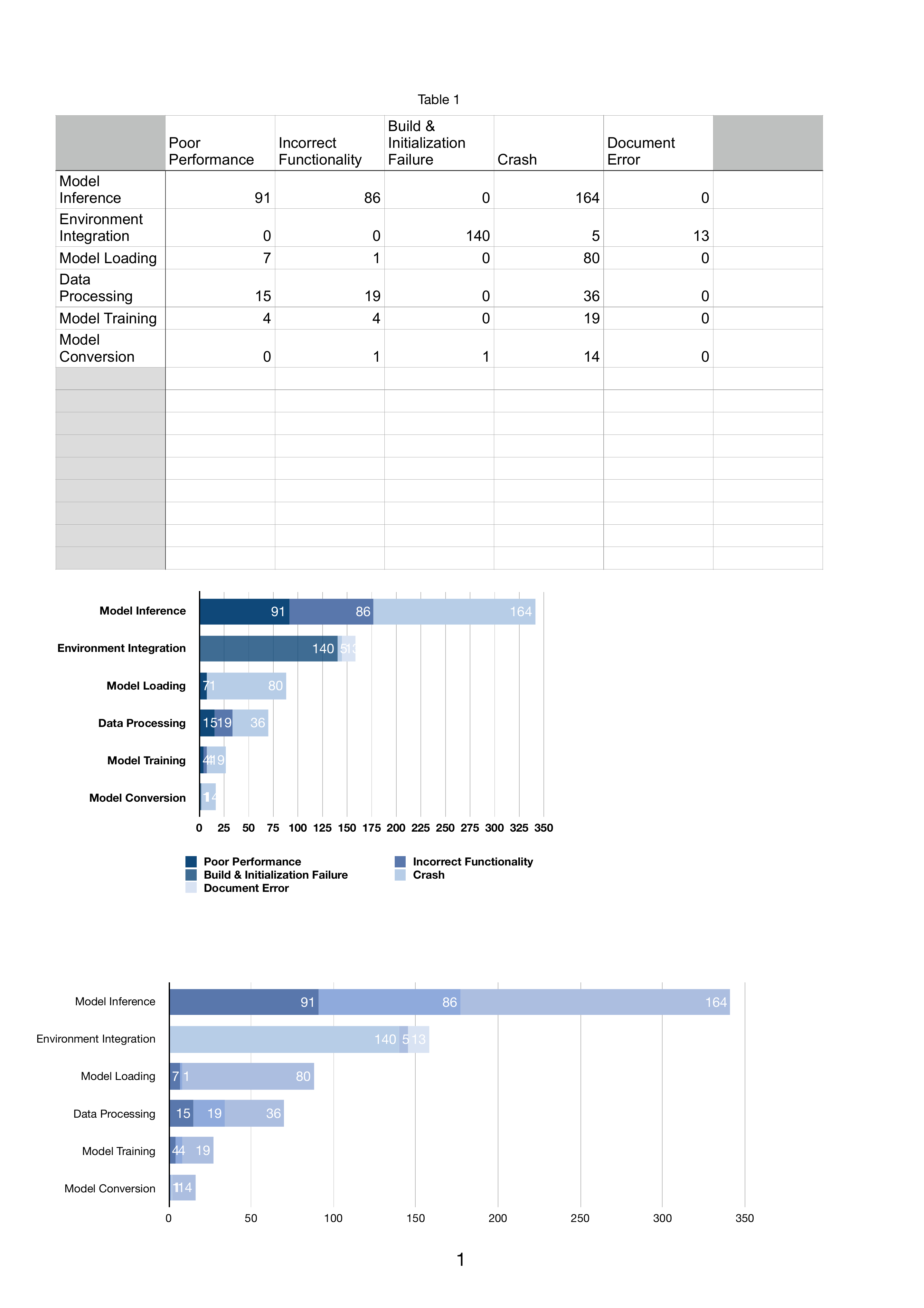}
    \caption{Symptom distribution in each stage}
    \label{fig:symptom-stage}
\end{figure}

We further investigate the symptom distributions in the 6 stages (See \Cref{fig:developing-stage}) to understand how the faults differ across different stages.
\Cref{fig:symptom-stage} shows the distribution in each stage where faults are exposed. As we can see, the symptoms at different stages are different. 
Particularly, \textit{Crash} and \textit{Poor Performance} are the common symptoms in all stages (except \textit{Environment Integration}), especially the \textit{Model Inference} stage.
Build \& Initialization Failure mainly appear in \textit{Environment Integration} stage.

In terms of stages, all faults are scattered throughout the life cycle of \target DL system, including model development (e.g., model training) and model deployment. \textit{Model Inference} and \textit{Environment Integration} are the most error-prone stages, accounting for 71.28\% of all faults. By contrast, \textit{Model Conversion} is the stage with the least errors.
This fault distribution is quite different from that on mobile DL applications~\cite{chen2021empirical}, e.g., most faults (i.e., 48.4\%) in mobile DL applications occur during the \textit{Model Conversion} stage. 
This is caused by the differences between JavaScript DL applications and mobile DL applications. 
For example, the models in mobile DL applications are usually pretrained on servers and then converted for mobile usage, while models in \target DL applications can be directly trained on browsers or servers with Node.js.
Besides, the complex and diverse runtime environment (e.g., different backends) in TensorFlow.js makes that more faults belong to Environment Integration, which is also different from native DL frameworks.

\vskip 1mm
\noindent \fbox{
	\parbox{0.95\linewidth}{	\textbf{Finding}: 
		Faults exposed at different stages present different symptoms. \textit{Crash} and \textit{Poor Performance} are the top two symptoms that go through the entire life cycle of \target DL systems. 
		\textit{Model Inference} and \textit{Environment Integration} are the most error-prone stages, covering 71.28\% of all faults, due to the complexity of the architecture of the \target DL system (multi-level and multi-component). }
}

%% file: 4-rootcause.tex
\section{Root Causes (RQ2)}\label{RootCause}
\subsection{Root Cause Classification Results \label{sec:root-cause-classification}}
\begin{figure*}
    \centering
    \includegraphics[width=0.72\paperwidth]{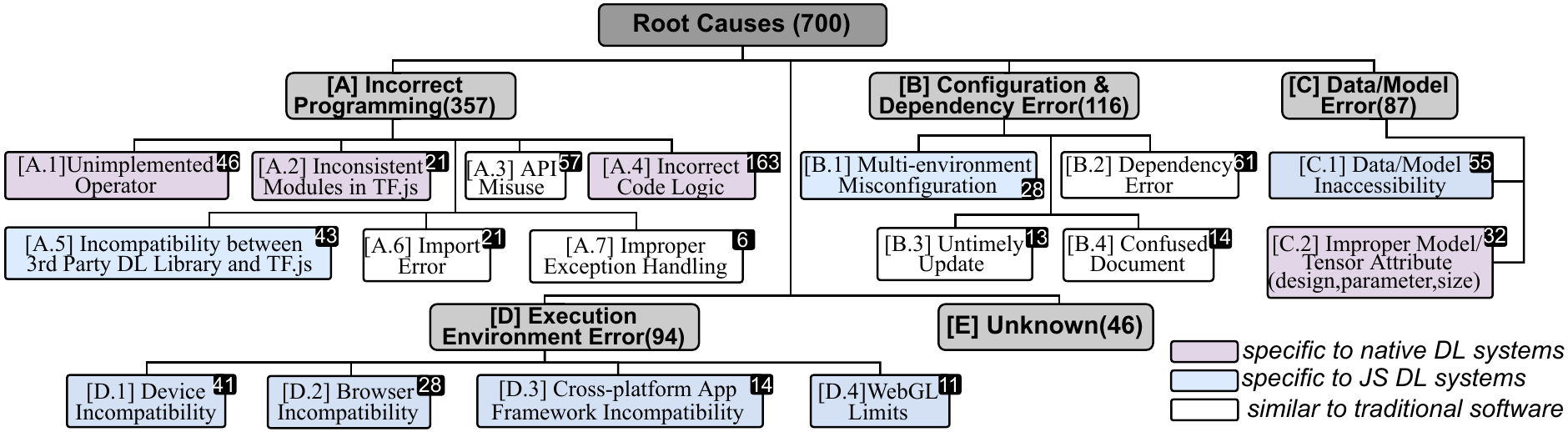}
    \caption{\revised{Root cause taxonomy of faults in \target DL systems
    }
    \label{fig:rootcause}}
\end{figure*}

The root cause taxonomies of the studied faults are shown in \Cref{fig:rootcause}, which is organized into 5 high-level categories (i.e., \textit{Incorrect Programming}, \textit{Execution Environment Error}, \textit{Configuration \& Dependency Error}, \textit{Data/Model Error} and \textit{Unknown}) and 17 inner categories.
The number of faults assigned to each category is in the top right corner.
Note that, there are 46 (6.57\%) faults that are difficult to analyze their root causes, and are therefore classified as \textit{Unknown}.
\revised{The blue and pink rectangles indicate root causes specific to JavaScript-based DL systems and native DL systems respectively, which will be further detailed in \Cref{sec:root cause difference}.}

\subsubsection{\textbf{Incorrect Programming (A)}\label{sec: Incorrect Programming}}
This category covers faults caused by program code, which is the most common category and accounts for 357 (51.00\%) of the faults. It contains 7 subcategories i.e., \textit{Unimplemented Operator}, \textit{Inconsistent Modules in TF.js}, \textit{Incorrect Code Logic}, \textit{Incompatibility between 3rd-party DL Library and TF.js}, \textit{API Misuse}, \textit{Import Error}, and \textit{Improper Exception Handling}. 

\textit{Incorrect Code Logic} causes the most faults, accounting for 23.29\% of all faults in this study, which can be divided into three subcategories based on the code functionality. 
\ding{172} Incorrect DL-specific algorithms due to incorrect implementation of DL-specific functions in TensorFlow.js (e.g., basic DL operators~\cite{issues5641}), 3rd-party DL libraries (e.g., face recognition algorithm~\cite{issues66}), and incorrect code logic in web applications (e.g.,~\cite{issues461}).
The most \textit{Incorrect Code Logic} faults (119 faults) belong to this subcategory.
\ding{173} Incorrect memory management algorithm due to improper memory management, accounting for 24 (3.43\%) of all faults. It mainly appears when explicit memory management is used on the WebGL backend.
For example~\cite{issues4378}, there is a memory leak in operator \texttt{tf.signal.stft}, because some intermediate tensors are not released in time. The TensorFlow.js vendors explained it as \textit{``Complex components cannot be released if there are multiple references on the components and those references are disposed before the complex tensor is disposed.''}
\ding{174} Poor environmental adaptability due to the incorrect/missing condition checking (i.e., \textsf{if-else} blocks) required to handle different environments (e.g., specific browsers). It accounts for 20 (2.86\%) of all faults. Such faults mainly occur in 3rd-party libraries and TensorFlow.js. 
For example~\cite{issues5334}, the application works well on Chrome and Edge but fails on Opera. The TensorFlow.js vendors explained that \textit{``In certain cases (e.g. in a webworker running in Opera), the window is not defined, which will fail the isMobile function. This PR adds a fallback check which uses navigator.userAgentData.mobile.''}

46 faults are caused by \textit{Unimplemented Operator}, i.e., the DL operators used in DL systems are not yet supported or implemented by TensorFlow.js. 
For example~\cite{issues5110}, the operation \texttt{tf.mod} is not supported by the Wasm backend of TensorFlow.js.
Besides, there are many modules in TensorFlow.js (e.g., the \texttt{tfjs-core} and \texttt{tfjs-tflite}), which cooperate with each other to complete various DL tasks and adapt to various environments.
Inconsistent implementations between these modules lead to 21 faults in this study \revised{(see \textit{A.2} in ~\Cref{fig:rootcause}}). 
For example~\cite{issues5700}, \texttt{tfjs-core} does not support tensor of type Int8Array provided by another module \texttt{tfjs-tflite}.

Apart from the faults caused by TensorFlow.js implementations, 43 faults are introduced by the 
flaws in 3rd-party DL libraries, i.e., \textit{A.5} in ~\Cref{fig:rootcause}. 
It refers to faults caused by the wrong TensorFlow.js versions, as the 3rd-party DL library requires the specific version of TensorFlow.js.
For example~\cite{issues794}, the library \texttt{face-api.js} executes based on outdated versions of TensorFlow.js (i.e., 2.x), resulting in an incompatibility error ``\texttt{t.toFloat} not being a function''. 

\noindent \fbox{
	\parbox{0.95\linewidth}{	\textbf{Finding}: \textit{Incorrect Programming} is the most common root cause category and covers 7 subcategories, accounting for 357 (51.00\%) of all faults. Among them, \textit{Unimplemented Operator}, \textit{Inconsistent Modules in TF.js}, and \textit{Incorrect Code Logic} are related to the implementations of TensorFlow.js. The most common subcategory is \textit{Incorrect Code Logic}, especially the incorrect implementation of DL-specific algorithms.}
}

The remaining 3 root causes are common in traditional software. 
Specifically, \textit{API Misuse} refers to the faults due to the users' misunderstanding of APIs, including missing or redundant calls to an API, wrong API names, and invalid API input/parameters (i.e., type/shape/value error). 
It contains 57 faults, accounting for 8.14\% of all faults. Our further analysis finds 52.63\% (30/57) of \textit{API Misuse} faults are due to invalid usage of inputs/parameters, indicating 
that in \target DL systems, developers are confused about the types and shapes of parameters/inputs supported by the APIs, especially for data types that can only be used in specific environments.
\textit{Import Error} refers to the faults (21 faults) caused by the missing/incorrect import of TensorFlow.js, and the import of multiple versions of TensorFlow.js at the same time. 
The remaining 6 faults are due to \textit{Improper Exception Handling}, including missing exceptions, suspicious exceptions, and confusing error messages.

\vskip 1mm
\noindent \fbox{
	\parbox{0.95\linewidth}{	\textbf{Finding}: For the root causes that are also common in traditional software, API Misuse accounts for a considerable amount, especially the invalid API input/parameters. }
}

\subsubsection{\textbf{Configuration \& Dependency Error (B)}}
116/700 (16.57\%) faults are caused by the incorrect configuration and dependencies, which is the second most common category. 
\revised{In particular, 28 \textit{Multi-environment Misconfiguration} faults are specific to \target DL systems. They are caused by incorrect bundler configurations that are used to ensure the same implementation of a \target DL system can be deployed on heterogeneous environments (e.g., browser and Node.js), regardless of the underlying hardware types (i.e., PC, smartphones, and wearable devices) and the operating systems (e.g., Windows, iOS, and Android).
For example, a fault is caused by not marking ``os'' as the external attribute in the bundler configuration for the browser target~\cite{issues4745}.}

74 faults are caused by dependency-related problems, of which 61 faults suffer from the missing/redundant dependency, the dependency version mismatch, and dependencies with security vulnerabilities~\cite{issues5492}
(see \textit{Dependency Error}); the remaining 13 faults are caused by the untimely updates of \textit{tensorflow.so}~\cite{issues5702} and npm packages (see \textit{Untimely Update}).
These dependency faults are closely related to the characteristics of TensorFlow.js, which relies on various libraries. For example, XNNPACK~\cite{XNNPACK} is a highly optimized library for floating-point neural network inference that can be used on ARM, WebAssembly, and x86 platforms.
Such complicated dependencies will inevitably introduce fragility during the configuration and runtime. 
The dependencies that are not updated in time or the relevant properties are not given correctly may cause serious problems in the entire system. 
Another 14 faults in this category are due to \textit{Confused Document} in TensorFlow.js. Although the number is small, it brings bad experiences to users.

\vskip 1mm
\noindent \fbox{
	\parbox{0.95\linewidth}{	\textbf{Finding}: \textit{Configuration \& Dependency Error} is the second most common root cause, covering 116 (16.57\%) of all faults. \textit{Multi-environment Misconfiguration} and \textit{Dependency Error} constitute two notable root causes, which are closely determined by the characteristics of TensorFlow.js (i.e., it depends on various libraries and can be used on multiple environments/platforms).}
}

\subsubsection{\textbf{Data/Model Error (C)}}
\revised{DL model and data introduce 87 (12.43\%) faults. 
Particularly, 55/87 faults are caused by the \textit{Data/model Inaccessibility}, due to 
1) the browser limitations, i.e., local data/model cannot be accessed because of the same-origin policy ~\cite{Same-origin-policy} that browsers follow;
2) the UI framework limitations, i.e., model is not placed in specified folders (e.g., public/asset) as required by the UI framework;
3) the incorrect model path or extension. If developers are unfamiliar with the features of browsers and UI frameworks, it is easy to introduce the inaccessibility of models or data. }

\revised{The remaining 32 faults are caused by \textit{Improper Model/Tensor Attribute}, including the poor model design (e.g., incorrect inference due to the poor quality of models provided by the 3rd-party DL libraries~\cite{issues16}), improper model parameter (e.g., long inference time due to the large input size~\cite{issues32}), and improper model size (e.g., ssd mobilenetv1 model cannot run on Android because it requires a lot of resources ~\cite{issues77}).
This shows that the quality of the pre-trained models provided in the 3rd-party DL libraries needs to be improved, and some models with large size cannot work well due to the limited computing power of the web platform.}

\vskip 1mm
\noindent \fbox{
	\parbox{0.95\linewidth}{	\textbf{Finding}: 63.22\% of \textit{Data/Model} faults are caused by the Data/Model Inaccessibility. Such faults are mainly related to the limitations of browsers and UI frameworks.}
}

\subsubsection{\textbf{Execution Environment Error (D)}}

As stated before, TensorFlow.js is designed to execute on various environments, such as different backends (e.g., WebGL and Wasm) for browsers, and cross-platform applications~\cite{TensorFlow.js}. 
In this study, 94/700 (13.43\%) faults are caused by imperfect support of TensorFlow.js for some hardware/software environments, which can be further divided into 4 subcategories, i.e., the \textit{Device Incompatibility}, \textit{Browser Incompatibility}, \textit{Cross-platform App Framework Incompatibility}, and \textit{WebGL Limits}.

\textit{Device Incompatibility} is the major environmental root cause, covering the largest number of faults (41/94 faults).
Such issues persist when TensorFlow.js executes on specific hardware (e.g., graphics card) and operating systems (e.g., Android). For example~\cite{issues299}, the DL system gives abnormal results
on devices equipped with Intel HD Graphics. 
Compared with native DL frameworks~\cite{chen2022toward}, such faults are more prominent in \target DL frameworks, indicating the implementation of DL backends specific to JavaScript usage needs to be improved in order to fit more diverse devices.

\textit{Browser Incompatibility} (i.e., PC browser and mobile browser) is another major cause, covering 28 faults. Such faults are reasonable because the DL inference on browsers is executed in JavaScript and relies on the browser engine for interpretation. 
There are also 14 faults caused by \textit{Cross-platform App Framework Incompatibility}. Specifically, TensorFlow.js can be integrated into mobile/desktop applications via cross-platform application frameworks (e.g., React Native). However, some of them are not compatible with the underlying libraries (e.g., \textit{expo-gl}) on which TensorFlow.js depends. 
Apart from the incompatibility factors mentioned above, the inherited limitations of WebGL (a set of JavaScript APIs which can be used for accelerating DL tasks) can also bring faults in some browser-based DL scenarios (14 faults). For example, the GUI is blocked due to the management mechanism of GPU resources in WebGL~\cite{issues5454}.

\vskip 1mm
\noindent \fbox{
	\parbox{0.95\linewidth}{	\textbf{Finding}: \textit{Execution Environment} is a common root cause that is quite specific to \target DL systems, due to the complex software/hardware environments they execute on. \textit{Device Incompatibility} and \textit{Browser Incompatibility} are top two environmental causes, accounting for 73.40\% of this category.}
}

\subsection{Root Cause Distribution}

\begin{figure}[t]
\centering
\includegraphics[width=0.33\paperwidth]{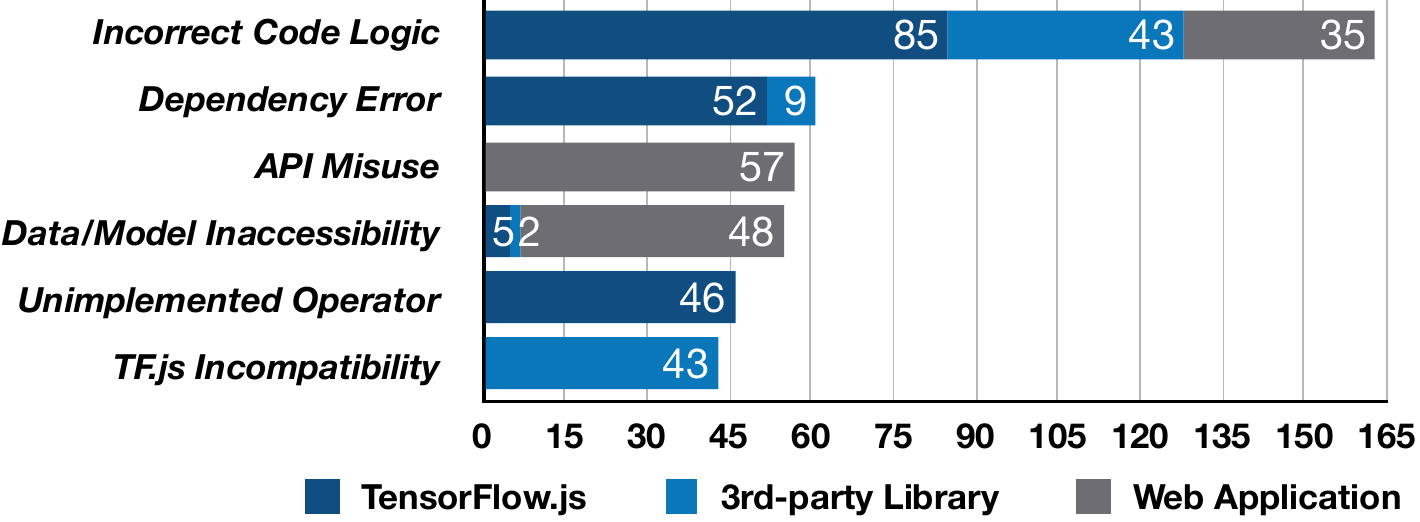}
\caption{The distributions of root causes on 3 levels of DL systems} \label{fig:rootcaues-distribution}
\end{figure} 

\subsubsection{\textbf{Distribution on the 3 levels of DL systems.}\label{sec:distribution-on-3level}}
We further analyze the distribution of root causes to understand how these faults present on the 3 levels of \target DL systems (i.e., Web application, 3rd-party DL library, and TensorFlow.js).
As shown in~\Cref{fig:rootcaues-distribution}, the top 6 common root causes (425/700 faults, 60.71\%) are considered due to the space limit. Note that the \textit{TF.js Incompatibility} stands for \textit{A.5} in ~\Cref{fig:rootcause}.
 
In terms of root cause, 
\textit{Incorrect Code Logic} is the most common type, which distributes over all of the 3 levels. Such faults should be a concern for all developers and researchers. The remaining root causes are distributed over a specific level.
For example, 85.25\% of \textit{Dependency Errors} and all of the \textit{TF.js Incompatibility} faults appear in TensorFlow.js and 3rd-party library, respectively.
In terms of system levels, most faults are caused by TensorFlow.js. 
Different levels present different fault distribution tendencies.
Specifically, faults on TensorFlow.js are closely related to low-level implementation, e.g., \textit{Incorrect Code Logic} and \textit{Unimplemented Operator}, indicating that TensorFlow.js is still at the early stages of development. The vendors need to enrich DL operators and check the libraries on which TensorFlow.js depends in time.
As a comparison, faults on 3rd-party libraries are mainly caused by \textit{TF.js Incompatibility}, and faults on web applications are more about the high-level usage of TensorFlow.js, e.g., the \textit{API Misuse} and \textit{Data/Model Inaccessibility}. In these cases, the 3rd-party library vendors need more effort to keep compatible with TensorFlow.js, and web application developers should carefully use DL-related APIs and handle model/data.

\subsubsection{\textbf{Distribution on the framework components.}\label{sec:distribution-on-components}}
\begin{figure}[t!]
	\centering
	\includegraphics[width=0.33\paperwidth]{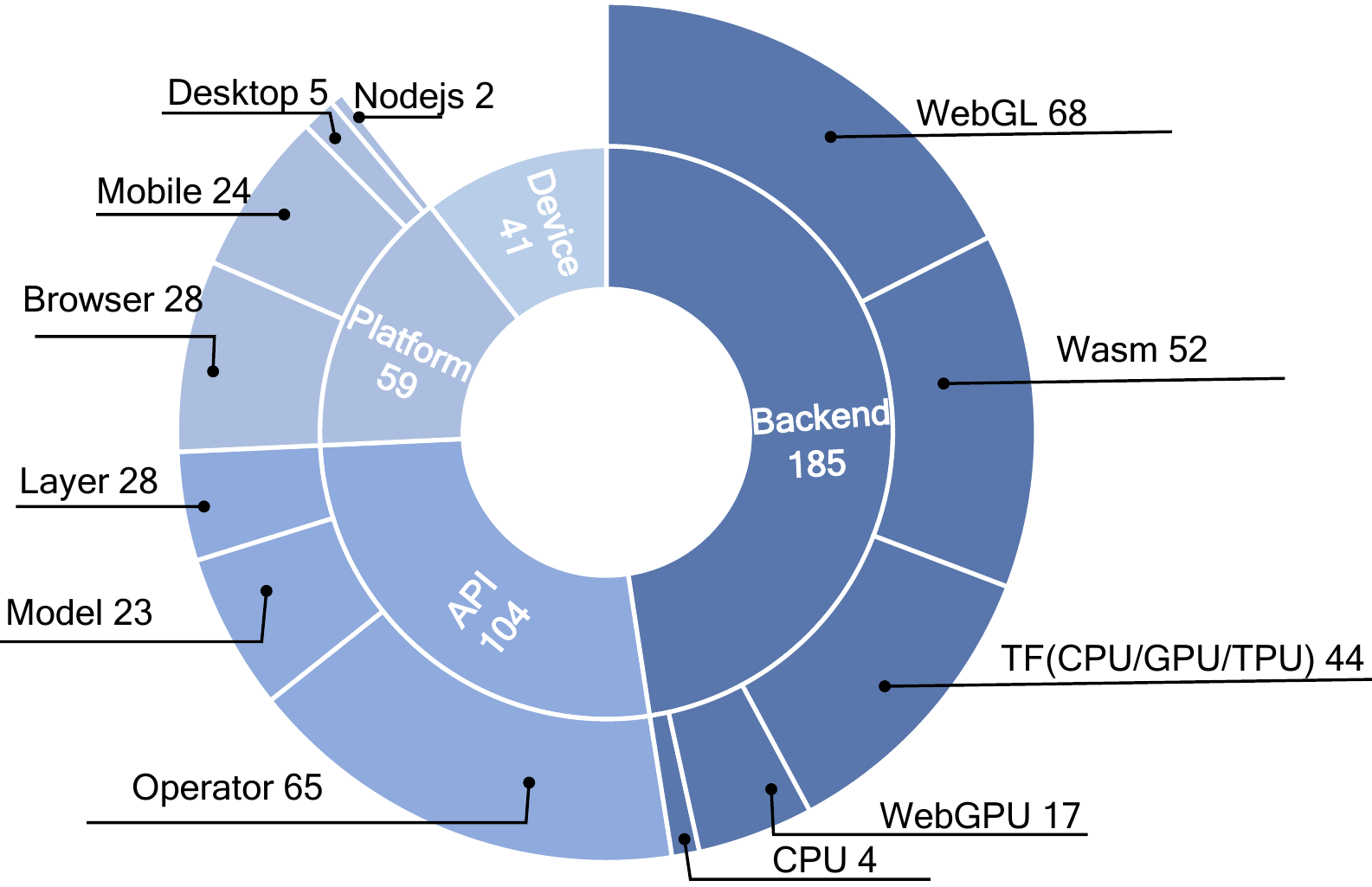}
	\caption{The distribution of faults across TensorFlow.js components}
	\label{fig:root-cause-components}
\end{figure}

TensorFlow.js contains 4 components (see~\Cref{fig:architecture-of-jsdl-system}), i.e., API, Platform, Backend, and Device.
To evaluate the quality of them, we further analyze the distribution of faults on each component, as shown in~\Cref{fig:root-cause-components}.
The inner layer represents the 4 components and corresponding faults number.
The outer layer represents the sub-component corresponding to the component (the inner node of the same color) and the number of errors caused by each sub-component.

As we can see, 389/700 (55.57\%) faults are introduced by the 4 TensorFlow.js components. 
DL-backend-related faults cover the most faults (i.e., 185/389 faults), of which the JavaScript-specific backends (i.e., WebGL, Wasm, WebGPU, and Pure-JS CPU) account for the majority of cases (141 faults), especially the WebGL (68/141) and Wasm (52/141). A large scale of faults brought by DL backends indicates that \target DL systems are still at dawn for the goal of conducting DL across multiple environments. In particular, JavaScipt-specific backend (e.g., WebGL) are more fragile than the native backend.
API component introduces the second most faults (104/389 faults), which consists of 3 levels of APIs, i.e., the operator-level, layer-level, and model level. Among them, the operator-level APIs bring the most faults (65 faults). Since existing testing techniques~\cite{pham2019cradle,chen2022toward,wang2020deep,guo2020audee} suffer from low operator coverage, we emphasize this is a challenge for detecting API errors on \target DL systems.
Regarding the Platform component, it causes 59 faults, which mainly occur on browsers and mobile applications.
The remaining 41 faults are due to device components.

\vskip 1mm
\noindent \fbox{
	\parbox{0.95\linewidth}{	\textbf{Finding}: In terms of the 3 system levels, the root causes present certain tendencies on different levels. \textit{Incorrect Code Logic} is the most common one that affects all of the 3 levels. In terms of the components in TensorFlow.js, most faults are caused by the DL backends, especially the JavaScript-specific backends (e.g., WebGL), which calls for new testing techniques and debugging methods for detecting such errors.}
}

%% file: 5-fixpattern.tex
\section{Fix Patterns (RQ3)}\label{FixPattern}

A fix pattern means that the subject fixes these faults by using them to modify the object.
\Cref{tab:fix_pattern} shows the 16 common fix patterns with 539 faults, which involve 4 major objects (i.e., Environment, Model, Data, and Program \& API) and 2 subjects (i.e., application developers and framework/3rd-party library vendors).
Note that we consider the subject to be the framework/3rd-party library vendors if the fault is fixed via PR in the framework/third-party library repository, otherwise it is the application developers.

\noindent\textbf{Environment.}
Modifying the environment configuration is the most common, which resolve 264 faults and is typically used by developers.
Specifically, changing versions of the TensorFlow.js, 3rd-party DL libraries, and compiler/installer (e.g., Typescript, Node.js, and NPM) can solve most faults (i.e., 99).
89 faults can be resolved by changing the browser/device, backend, and values of environment variables.
Besides, 13 import confusion in the program can be fixed by adding/removing imports and changing import statements.
Note that the pattern of modifying the dependency configuration includes adding dependencies, removing dependencies, replacing mismatched dependencies, and modifying related configuration options, which can be used by both developers and vendors.

\begin{table}[t!]
\centering
\scriptsize
\caption{Fix patterns of the faults}
\label{tab:fix_pattern}
\begin{tabular}{@{}l|r|r|c@{}}
\toprule
\multicolumn{1}{l|}{\textbf{Object}} & \textbf{Subject} & \textbf{Fix Pattern} & \textbf{\#} \\ \midrule
\multirow{6}{*}{\textbf{Environment}} & Developer & Changing version & 99 \\
 & Developer, Vendor & Modifying dependency configuration & 63 \\
 & Developer & Changing device/browser & 39\\ 
 & Developer & Changing backend & 31\\
 & Developer & Modifying the value of environment variable & 19 \\ 
 & Developer & Fix import confusion in program & 13 \\
 \midrule
 \multirow{2}{*}{\textbf{Model}} & Developer & Modifying model file path/extension & 37\\
 & Developer & Changing model & 32 \\ \midrule
\multirow{2}{*}{\textbf{Data}} &Developer & Add data processing & 27 \\
 & Developer, Vendor & Replace data shape/type & 27 \\ \midrule
\multirow{6}{*}{\textbf{Program \& API}} & Vendor & Add unsupported operator & 40 \\
 & Developer, Vendor & Replace API with another effective one & 29 \\
 & Developer & Modify API parameter usage & 25 \\
 & Developer, Vendor & Fix environment adaptability & 20 \\
 & Developer, Vendor & Adjust API invocation sequence & 19 \\
 & Developer, Vendor & Add API usage for memory management & 19 \\ \midrule
\textbf{Total} & -&- & \textbf{539} \\ \bottomrule
\end{tabular}
\end{table}

\smallskip
\noindent\textbf{Model.}
Developers fix 69 faults by modifying the models.
Specifically, modifying the model file path/extensions solves 37 faults, which are mainly model inaccessibility faults.
Another 32 faults are fixed by model reconstruction, including retraining the model, reconverting the model, and replace with another similar model.

\noindent\textbf{Data.}
There are two patterns that act on the data, including adding data processing and replacing data shape/types.
Specifically, developers fix 27 faults by adding preprocessing of inputs and post-processing of model predictions.
Another 27 faults are fixed by developers and vendors by replacing the shape/type of related data.

\noindent\textbf{Program \& API.}
152 faults are resolved by modifying the related programs and API usage.
For this object, there are 6 fix patterns, 1 of which is typically used by the vendors (i.e., add unsupported operator), 1 by the developer (i.e., modify API parameter usage), and the remaining 4 can be used by both the developers and the vendors.
Specifically, vendors solve 40 faults (i.e., unimplemented operators) by adding unsupported operators.
Developers solve 25 API Misuse faults by modifying the usage of the API parameters.
The remaining 4 patterns resolve 87 faults that can be introduced by any of TensorFlow.js, 3rd-party library, and web application.

Note that, the fix pattern applied to one object can also repair faults caused by other objects.
For example, for the pattern of changing versions in the environment, developers can also use it to resolve incorrect program logic errors in TensorFlow.js in addition to faults caused by the environment, because vendors fix bugs inside TensorFlow.js and update the TensorFlow.js version frequently.
Additionally, faults caused by the same root cause can be fixed by different patterns.
For example, for an error caused by the Wasm backend not supporting a certain operator, the developer can bypass the faults by changing the backend into WebGL backend, however, the vendor can add support for the operator to solve the faults.
Although both methods can solve the faults, adding support for the operator is the most direct and effective method.

\vskip 1mm
\noindent \fbox{
	\parbox{0.95\linewidth}{		\textbf{Finding}: We summarize 16 common fix patterns based on 2 subjects and 4 objects. \revised{Modifying the environment is the most common pattern, especially changing versions of the TensorFlow.js, 3rd-party DL libraries, and compiler/installer.} We find that faults caused by the same root cause can be fixed by different patterns in practice. }
}

%% file: 6-difference.tex
\section{Differences from native DL systems (RQ4)}
In this section, we discuss the differences between the taxonomies proposed in this study and that proposed by previous work on other systems from 2 aspects: the symptom and the root cause.
\subsection{Differences Based on Symptom \label{sec:symptom difference}}
\revised{
As highlighted by the pink rectangles in~\Cref{fig:symptom-classification},
5 symptoms (e.g., \textit{DL Operator Exception}) involving the characteristics of DL are shared with the existing symptom taxonomies for native DL (e.g., TensorFlow and TensorFlow Lite) faults~\cite{chen2021empirical,chen2022toward,gu2021demystifying,humbatova2020taxonomy}. 
Despite all this, their symptoms on different DL frameworks are not exactly the same. For example, we found some cases where a basic DL operator/training argument is problematic in TensorFlow.js, but it is supported in TensorFlow (see~\cite{issue4852}).
This shows that TensorFlow.js needs to be aligned with the native framework in providing complete implementations of DL operators and training parameters.}

Moreover, 8 symptoms are closely related to the characteristics of \target DL systems, as shown by the blue rectangles in ~\Cref{fig:symptom-classification}, covering 35\% faults. 
In Particular, the \textit{Multi-backend Initialization Failure} and \textit{Inconsistency between Backends/Platforms/Device} are determined by the characteristics of TensorFlow.js, which provide several parallel DL backends (e.g., WebGL and Wasm) specific to the different JavaScript execution environment. 
Similarly, The \textit{Fetch Failure} and \textit{Browser\&Device Error} are determined by the fact that \target DL systems mainly run on heterogenous browsers, including both PC browsers and mobile browsers. 
Such new symptoms cover more than one-third of the faults in this study, indicating that the quality of \target DL systems deserves a comprehensive investigation.

The remaining symptoms are similar to traditional software, as shown by the white rectangles in~\Cref{fig:symptom-classification}.
The performance faults account for a large proportion in this study.
Considering the low performance of JavaScript DOM manipulation and the explicit memory management of the WebGL backend, \target DL systems are more prone to performance issues, which should arouse developers' attention and be analyzed exclusively in the future.

\subsection{Differences Based on Root Cause \label{sec:root cause difference}}
\target DL systems experience some common issues as is for native DL systems, which have been extensively studied in prior work~\cite{chen2022toward,xiao2018security,jia2021symptoms}. As shown by the pink rectangles in~\Cref{fig:rootcause}, these issues are primarily caused by incorrect code logic (163 faults), unimplemented operators (46 faults), improper model/tensor attribute (32 faults), and inconsistent implementations between different modules in TensorFlow.js (21 faults). Although the aforementioned problems are expected, it is crucial to stress that a total of 262 faults (over 37\%) brought on by typical DL-specific problems illustrate the incompleteness of \target DL systems in supporting core DL functionalities.

Besides, we summarized 220 faults (31.4\% of all faults) whose root causes are specific to \target DL systems, covering 7 major subcategories. 
As shown by the blue rectangles in~\Cref{fig:rootcause}, 126 faults are caused by the incompatibility problem.
These compatibility issues can be summarized into two levels:
1) the 3rd-party DL library level, i.e., the incompatibility between TensorFlow.js and the 3rd-party DL libraries that wraps TensorFlow.js (43 faults);
2) the environment level, i.e., the incompatibility of TensorFlow.js with various execution environments, including the devices (41 faults), browsers (28 faults), and cross-platform applications (14 faults).
Notably, another 28 faults are caused by the unique multi-environment characteristic of \target DL systems. Compared with other single-environment DL systems (e.g., mobile DL apps), the \target DL systems are designed to run in a variety of environments that rely on different configurations (e.g., browsers and Nodejs).
In addition, 55 faults are mainly caused by the limitations of the browser and UI framework
(e.g., Local data/models inaccessible due to the same-origin policy~\cite{Same-origin-policy} in browsers),
and 11 faults are caused by WebGL limitations (e.g., GUI block due to the WebGL management mechanism of GPU resources). Such new root causes introduce more than 30\% faults, suggesting the differences between the JavaScript-based systems and others.

The remaining categories in~\Cref{fig:rootcause} are common causes in traditional software, as shown by the white rectangles. 
The dependency-related error (61 faults) and API misuse (57 faults) are two leading factors.
Particularly, as for \textit{API Misuse}, 52.6\% API-misuse cases are due to the invalid inputs/parameters in \target DL systems, which is quite different from traditional software~\cite{amann2016mubench,amann2018systematic} and even native DL framework~\cite{chen2022toward}, where the API missing/redundancy, and incorrect API names are uppermost cases. 
Therefore, for the \target DL systems, some existing problems in traditional software are still worth exploring in combination with the characteristics of these systems.

\vskip 1mm
\noindent \fbox{
	\parbox{0.95\linewidth}{		\textbf{Finding}: 
		8 symptoms (35\% of all faults) in our study are specific to the \target DL systems, suggesting that the quality of JavaScript-based DL systems deserves a comprehensive investigation.
		7 root causes (nearly one-third of all faults) in our study are specific to the \target DL systems. Particularly, the incompatibility issues are prominent (18\%) in TensorFlow.js usage. More efforts are needed for TensorFlow.js vendors to adapt to more environments. }
}

%% file: 7-discussion.tex
\section{Discussion}\label{sec:discussion}
\subsection{Implications}
\subsubsection{\textbf{For Application Developers.}}

\textit{Data/Model Inaccessibility} and \textit{API Misuse} are major fault causes for the \target DL applications, as shown in \cref{sec:distribution-on-3level}. 
Therefore, we conclude some tips: \textbf{1)} Avoid data/model inaccessibility. Since data and models are the foundation of DL applications, developers should first ensure their accessibility.
Depending on the causes of such issues, we recommend that developers avoid such errors by \revised{\ding{172} putting the model on the Internet and using TensorFlow.js to load the online model via the URL to bypass the limitation that the browsers cannot directly access local files;}
\ding{173} checking if the model is placed in the folder required by the UI frameworks (e.g., \textit{Angular});
\ding{174} checking if the model path and extension are correct.
\textbf{2)} Use API carefully. 
Quite a few faults are caused by the confusion on the API parameters/inputs in TensorFlow.js. Therefore, developers should use the APIs carefully, such as understanding the usage of the API based on the official documentation before programming.

\subsubsection{\textbf{For 3rd-party DL Library Developers}.}
\textbf{1)} Improve the environmental adaptability.
\Cref{sec: Incorrect Programming} reveals that faults caused by poor environmental adaptability mainly appear in 3rd-party DL libraries. Developers are expected to conduct cross-platform testing before releasing a library, ensuring the library adapts to any platform, especially the browsers and Node.js.
\textbf{2)} Enhance compatibility with TensorFlow.js. 
Only a specific version of TensorFlow.js can work with 3rd-party DL libraries, which brings great inconvenience to users. 
Such issues should arouse the attention of developers.

\subsubsection{\textbf{For Framework Developers.}}
\revised{\textbf{1)} Enhanced testing of the implementations of DL backends specific to JavaScript (e.g., WebGL).
JavaScript-specific DL backends present more faults compared to native DL backends (see ~\Cref{sec:distribution-on-components}). Such faults should be noted and detected promptly.}
\textbf{2)} Do more unit testing. \textit{Incorrect Code Logic} is the most common root cause (see \cref{sec:distribution-on-3level}).
It causes various fault symptoms and is difficult to locate, so developers should focus on detecting such faults before releasing a new version to ensure that the implementation logic of each module is correct.
\textbf{3)} Check and update dependencies in time. Many faults in TensorFlow.js are related to the libraries on which it depends, especially the libraries with real vulnerabilities. To avoid these errors, we recommend that developers check and update dependencies in time.
\textbf{4)} Expand DL operators. Many operators have not been supported by TensorFlow.js, developers should support as many operators as possible to align with the mature native frameworks (e.g., TensorFlow).

\subsubsection{\textbf{For Researchers}.}
\textbf{1)} Call for testing techniques for performance faults.
In terms of symptoms, poor performance faults are prominent in \target DL systems and the reasons for such faults are difficult to analyze. Therefore, testing and debugging techniques for such faults are desired.
\textbf{2)} Focus on testing the framework.
For the 3 levels in the \target DL system, framework introduces the most faults. Particularly, JavaScript-specific DL backends present more faults compared to native DL backends, but there is currently no testing techniques for such errors. Thus, how to design effective testing methods according to the characteristics of \target DL framework is a challenge for future research.

\subsection{Threats to validity}

The external threat to validity lies in the dataset.
First, the selection of our study subjects (i.e., TensorFlow.js, 3rd-party DL libraries, and web applications) may be biased. To mitigate this threat, we choose the most popular and representative \target DL framework (i.e., TensorFlow.js) as a base. The selected 3rd-party DL libraries and web applications are all built on TensorFlow.js. \revised{The findings based on TensorFlow.js-related systems can be largely applied to other \target DL frameworks (e.g., Paddle.js~\cite{Paddlejs} and WebDNN~\cite{webdnn}) as 
most of them can run on DL backends (e.g. WebGL) specific to JavaScript like TensorFlow.js.}
Second, we identify relevant GitHub repositories and issues related to TensorFlow.js based on keyword matching. Some candidates may be ignored due to the predefined keywords, which would introduce biased in the data construction. To mitigate such a threat, we follow the previous work~\cite{humbatova2020taxonomy} to carefully select effective keywords to ensure most of the relevant repositories and issues can be identified.
The internal threat to validity lies in our manual labeling process.
To minimize the subjectivity of researchers, two authors conducted the labeling process independently, and another arbitrator with 3-year DL development experience helps to reach an agreement through discussions.
Moreover, we leveraged Cohen’s Kappa coefficient to measure the inter-rater agreement of independent labeling.
The high kappa value indicates a high agreement between researchers.

%% file: 8-related-work.tex
\section{Related Work}\label{sec:related-work}

A number of empirical studies have emerged recently on analyzing the bugs relevant to DL/ML frameworks.
Thung et al.~\cite{thung2012empirical} first targeted ML systems (i.e., Apache Mahout, Apache Lucene, and Apache OpenNLP) and analyzed 500 bug reports. They focused on the frequencies, types, severity\&impact, fixing effort\&duration of these bugs.
Sun et al.~\cite{sun2017empirical} focused on ML frameworks (i.e., Scikitlearn, Paddle, and Caffe) and manually analyzed 329 real bugs to study the bug types and bug evolution.
Several studies focused on the bugs in DL frameworks and applications, which generally collected real faults from Stack Overflow and GitHub, and applied taxonomic methods for fault summarizing.
Specifically, \cite{humbatova2020taxonomy,islam2019comprehensive,zhang2018empirical,harzevili2022characterizing,islam2020repairing,ma2018deepmutation} studied the bugs symptoms, root causes, and effects of DL applications under PC platform, which rely on popular native DL frameworks (e.g., TensorFlow, Keras, and PyTorch). 
\cite{jia2020empirical,jia2021symptoms,du2020fault} analyzed the implementation bugs of TensorFlow itself in terms of the symptom, root cause, and fix pattern. Chen et al.~\cite{chen2022toward} extended to four DL frameworks (i.e. TensorFlow, PyTorch, MXNet, and DL4J), and analyzed the current testing status of DL frameworks.
Moreover, there are also some empirical studies focusing on specific bug types. For example, 
Gu et al.~\cite{gu2021demystifying} studied the training issues of developers in DL software.
Cao et al.~\cite{cao2021characterizing} characterized the performance bugs in DL systems. 
Tambon et al.~\cite{tambon2021silent} studied the silent bugs existed in DL frameworks.
Deploying DL techniques onto mobile platforms has currently become another trend. 
Chen et al.~\cite{chen2020comprehensive} built a taxonomy of specific challenges that developers encounter during the deployment of DL software. They further studied the deployment faults of mobile DL applications in terms of the symptoms and fix patterns~\cite{chen2021empirical}.  

The aforementioned studies target the DL systems on the PC or mobile platforms, which are built on top of native DL frameworks (e.g., TensorFlow and TensorFlow Lite). 
Different from them, firstly, we target the \target DL systems built on TensorFlow.js, which is totally different from native frameworks in terms of the implementations of DL backends and the execution environments. 
Secondly, previous studies analyzed the faults on a specific level (i.e., framework-level or application-level), while this study analyzes the faults over a 3-level architecture of \target DL systems, including the web applications, 3rd-party DL libraries, and TensorFlow.js. 
Thirdly, different from the existing studies~\cite{chen2020comprehensive,chen2021empirical} that are related to the deployment faults of DL models on mobile devices, we mainly focus on the faults related to both the development (e.g., model training) and deployment of DL models on multi-environments (i.e., browsers, Node.js, and cross-platform apps). Apart from the symptom and fix pattern analyzed in~\cite{chen2021empirical}, we also classify the root cause in detail and analyze the fault distribution on the 3 levels of \target DL system and 4 major TensorFlow.js components.
We further detail the different features of fault symptoms and root causes between other DL systems~\cite{chen2021empirical} and \target DL system in~\Cref{sec:Symptom}.

Moreover, various \target DL frameworks have been released to enable DL tasks on browsers. 
To understand how well these frameworks behave in practice,
Ma et al.~\cite{ma2019moving} measured the performance gap of 7 JavaScript-based frameworks when running different DL tasks on Chrome.
Guo et al.~\cite{guo2019empirical} aimed at the DL software deployment across different platforms, and investigated the performance gap when the trained models are migrated from the PC platform to mobile devices and Web browsers.
Instead of focusing on the performance problems, we focus on the characteristics of faults in TensorFlow.js, 3rd-party DL libraries, and the web applications built on top of TensorFlow.js.

%% file: 9-conclusion.tex
\section{Conclusion}\label{sec:conclusion}
In this work, we conducted the first comprehensive study on faults in \target DL systems by manually inspecting 700 related faults from 3 levels of GitHub repositories (i.e., TensorFlow.js, 3rd-party DL libraries wrapping TensorFlow.js, and applications based on TensorFlow.js). We constructed taxonomies for fault symptoms, root causes, and fix patterns, respectively. Besides, we also analyzed the distribution of symptoms from the 6 stages involved in the lifecycle of \target DL systems and analyzed the distribution of root causes based on the 3 levels in \target DL systems and the 4 components of the TensorFlow.js. Additionally, we highlighted the different fault features between \target DL systems and native DL systems. The symptoms, root causes, and fix patterns discovered by our study can be adopted to facilitate fault fix in \target DL systems. Finally, we discussed the implications for different stakeholders based on our findings.

%% file: 10-ack.tex
This work was partly supported by the National Natural Science Foundation of China (No. 62102284, 61872262), 
the Ministry of Education, Singapore under its Academic Research Fund Tier 1 (21-SIS-SMU-033), the National Research Foundation, Singapore under its the AI Singapore Programme (AISG2-RP-2020-019), the National Research Foundation, Prime Ministers Office, Singapore under its National Cybersecurity R\&D Program (Award No. NRF2018NCR-NCR005-0001), NRF Investigatorship NRF-NRFI06-2020-0001. 